\documentclass[pra,twocolumn,aps,showpacs]{revtex4-1}
\usepackage{graphicx}
\usepackage{color}

\begin{document}
\title{Phase transition to spatial Bloch-like oscillation in
squeezed photonic lattices}
\author{M. Khazaei Nezhad$^{1}$, A. R. Bahrampour$^{1}$, M. Golshani$^{1}$, 
S. M. Mahdavi$^{1,2}$ and  A. Langari$^{1,3}$}
\affiliation{$\ ^{1}$Department of Physics, Sharif University of Technology, Tehran 11155-9161, Iran\\
$\ ^{2}$Institute for Nanoscience and Nanotechnology, Sharif University of
Technology, Tehran , Iran \\
$\ ^{3}$ Center of excellence in Complex Systems and Condensed Matter
(CSCM), Sharif
University of Technology, Tehran 1458889694, Iran
}

\date{\today}

\begin{abstract}
We propose an exactly solvable waveguide lattice incorporating inhomogeneous
coupling coefficient. 
This structure provides a classical analogues 
to the squeezed number and squeezed coherent intensity distribution in quantum 
optics where the propagation length plays the role of squeezed amplitude. 
The intensity pattern is obtained in a closed form for an arbitrary
distribution of the initial beam profile. 
We have also investigated the phase transition to the spatial Bloch-like
oscillations by 
adding a linear gradient to the  propagation constant of each waveguides 
($ \alpha $). Our analytical results show that the Bloch-like oscillations
appear above a critical value for the linear gradient of
propagation constant ($ \alpha > \alpha_{c} $). 
The phase transition (in the propagation properties
of the waveguide) 
is a result of competition
%This critical behavior can be interpreted as interplay 
between discrete and Bragg diffraction. Moreover, the light intensity decay
algebraically along each waveguide at the critical point while it falls off
exponentially below the critical point ($ \alpha < \alpha_{c} $).
\end{abstract}

%\pacs{42.25.Dd, 42.65.Wi, 42.79.Gn, 72.15.Rn, 73.20.Fz}

\maketitle

%{\em Introduction.-}
\section{Introduction}

Waveguide lattices provide an inexpensive experimental tool to study some 
physical phenomena in several branches of physics such as condensed matter, 
quantum optics, atomic and molecular physics \cite{Repkiv,Replon,RepLed,Nat2003,DL1986,Nat2009}.
Nowadays, these lattices can be realized by several methods such as 
optical induced technique, lithographic pattern and laser writing methods \cite{Nat2007, PRL2008}. 
In the weak coupling regime, the coupling strengths between adjacent
 waveguides can be adjusted by appropriate change in the distance between guides \cite{Repkiv}. 
Recently an exact solvable Glauber-Fock photonic lattice has been proposed in Refs.
\cite{Fock2010, FockPRL, Fock2012,FockPRA}, in which the coupling coefficients are 
inhomogeneous and proportional to the square root of waveguide
labels (assuming an incremental sequential labels). 
These waveguide arrays provide an experimental tool to investigate
 interesting phenomena such as quantum random walk,
 photon bunching and anti bunching\cite{FockPRL, FockPRA}. 

In the context of condensed matter physics, the energy levels of a super lattice form the
Wannier-Stark ladders \cite{Bloch1928,DL1986} when a constant external force is
 applied on its electrons. The extended Bloch wave function of electrons
 is converted to the localized Wannier states \cite{Bloch1928, DL1986}.
Moreover, the electrons show a periodic motion in these localized states,
 known as Bloch oscillation. Similar periodic motions have been observed
 for the light propagation in periodic waveguide arrays \cite{Blockcurved,
BlockTemp,Nonlin bloch, Blochcorr.,Bloch2d2006}. 
In such arrays the propagation direction plays the role of
time, therefore the periodic motion is known as the spatial-Bloch 
oscillations \cite{Blockcurved,BlockTemp,Nonlin bloch, Blochcorr.,Bloch2d2006}. 
The effect of a constant external force can be simulated by exerting a linear transverse gradient 
on the propagation constants of waveguide array \cite{Blockcurved,BlockTemp,Nonlin bloch,
 Blochcorr.,Bloch2d2006}. To this end, the transverse temperature gradient in
 thermo optical waveguides, transverse current in photorefractive waveguides 
or a fixed curvature on waveguide array during the fabrication process can be 
implemented \cite{Blockcurved,BlockTemp,Nonlin bloch, Blochcorr.,Bloch2d2006}. 

In this paper, we propose a new type of exactly solvable semi-infinite
optical waveguide lattice with inhomogeneous coupling coefficients. This array
provides the classical analogues to the squeezed number and squeezed coherent intensity 
distribution in which the squeezed amplitude is proportional to the
propagation length \cite{Squeeze1989, Knight, Walls, Puri}.
However, this is a classical simulation of the probability distribution of
squeezed state in the Fock space. 
Let consider a cross section of the waveguide lattice as shown in 
Fig.\ref{Fig1} where each waveguide is corresponding to a state 
($|n>, n=0, \dots, \infty$)
in Fock space, i.e. $|n=0>$ represents the vacuum state. The occupation
probability of each Fock state is given by 
the light intesity of the corresponding waveguide. The propagation of light
within each waveguide (along z-direction) corresponds to the time evolution
of the occupation probability of a Fock state.
The squeezed number intensity distribution can be simulated classically by
injecting a 
light beam in a single guide, while the Poisson distribution is applied to obtain
the squeezed coherent intensity distribution.

We have also investigated the spatial Bloch oscillation in the proposed squeezed array. 
Surprisingly, we obtain the critical value of the strength of linear gradient index to observe the 
Bloch-like oscillation ($ \alpha \geq\alpha_{c} $). We discuss the long
distance
behavior of light intensity along waveguides which shows three different behaviors,
namely: exponential decay for $\alpha < \alpha_c$, algebraic decay at $\alpha = \alpha_c$
and Bloch oscillation for $\alpha > \alpha_c$.
This oscillation comes 
from the interplay between discrete diffraction and Bragg diffraction. 
Here, both the coupling 
coefficients and propagation constants are approximately proportional to the
waveguide labels ($n$)
which make a competition between discrete diffraction and Bragg diffraction.
Such a critical value does not exist in the Fock-Gluaber and 
homogeneous lattices where the couplings are proportional to $\sqrt{n}$
and constant, respectively,
while the propagation constants are proportional to $n$ which leads to
the dominant Bragg diffraction
and Bloch oscillation.

The paper is organized in four sections. Sec. \ref{model} is devoted to the theoretical 
model and the exact solution which has been presented in our work.
In Sec. \ref{sbo} , the spatial Bloch-like oscillation in the proposed waveguide lattices 
is investigated. Finally, we conclude and summarize our results in Sec. \ref{summary}. 

\section{Theoretical Model \label{model}}

\begin{figure}[t!]
\includegraphics[angle=0, width=1.0\linewidth]{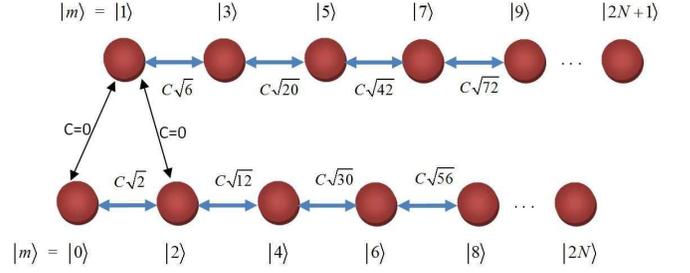}
\caption{(color online)
Cross-section of the squeezed array of optical waveguides.}
\label{Fig1}
\end{figure}

The non-degenerate squeezed operators depend on the 
%According to the dependence of non-degenerate squeezed operators on the 
square of annihilation and creation operators, which prohibits the squeezed
operators to be coupled
 to the even and odd Fock states, simultaneously. Therefore, to simulate the
 squeezed states, as shown in Fig. \ref{Fig1}, two different decoupled linear
array of waveguides are taken into account. 
The upper waveguide array is labeled by
 odd numbers while the lower array is labeled by even ones.
The distance between the odd and even arrays of waveguides is large enough
 to decouple the odd and even waveguides while the even-even and
 odd-odd nearest neighbor waveguides are coupled with nonzero coefficients. 
Although only one array (upper or lower) would be enough to simulate the 
vacuum squeezed and number squeezed states, 
both of upper and
lower arrays are necessary to be taken into account for simulating 
the coherent squeezed states, classically.
 
The slowly varying envelope approximation (SVEA) is implemented
 to write the light propagation equations in two 
linear waveguide arrays of Fig. \ref{Fig1}.
In this approximation the appropriate equations for light propagation 
in each waveguide is reduced to the common tight-binding (TB) model
\cite{Repkiv} as follows:
 \begin{eqnarray}
  \label{eq1}
  i\frac{d E_n(z)}{dz}+K_n E_n+C_{n} E_{n-2}
   	+C_{n+2} E_{n+2} =0,     	      
\end{eqnarray}
where, $E_n$ and $K_{n}$ are amplitude of the electric field and the propagation
constant of the $n^{th}$ waveguide, respectively. In this section, we consider
 identical waveguides and assume $K_{n}=K_{0}$. Moreover, $C_{n}$ is the 
coupling coefficient between the $n^{th}$ waveguide and it's preceding neighbor, $ n-2 $.
In the weak coupling regime, the coupling coefficients depend on the distance between
neighboring waveguides such that $C_{n} = C_{1} \exp[-\frac{d_{n}-d_{1}}{\kappa}]$
 where $ C_1 $ and $ d_{1} $ are coupling coefficient and distance 
between the first coupled waveguides in the upper (odd labeled) or 
lower (even labeled) arrays, respectively. 
$ \kappa $ is a free parameter which is determined from coupled mode
theory or experimental data \cite{Fock2010, FockPRL,Yariv}.
 In our model, we assume that the coupling coefficients between 
even-even or odd-odd waveguide arrays are determined by $ C_{n} = C_{1} \sqrt{n(n-1)} $ ($ n>1 $). 
It can be realized if we manipulate such that the distance between waveguides
are given by $ d_{n} = d_{1}-\frac{\kappa}{2} \ln[n(n-1)] $. 

The new variables $ Z = C_{1}z $ and $ E_{n}(z) = \Psi_n(Z)\exp(iK_0Z) $, 
transform equation (\ref{eq1}) to the following dimensionless form:
\begin{eqnarray}
\label{eq2}
i\frac{d \Psi_n}{dZ}(Z)+\sqrt{n(n-1)}\Psi_{n-2}(Z) \hspace*{2.8cm}
\nonumber \\
\hspace*{3mm}  +\sqrt{(n+1)(n+2)}\Psi_{n+2}(Z)=0,\hspace*{3mm} n =0,1,2,...	      
\end{eqnarray}

%%%%----------figure1------------------------------------------------
%%%%------------------------------------------------------------------

In order to find the solution of equation (\ref{eq2}), the following operator relation is defined:
 \begin{eqnarray}
  \label{eq3}
  i\frac{d \Phi}{dZ}(Z)=-(\widehat{a}^{2}+{\widehat{a}^{\dagger}{}}^{2})\Phi(Z), 
\end{eqnarray}
in which $ \Phi (Z) \equiv \sum \Psi_{m}(Z)\vert m\rangle $, where $ \vert m\rangle $
represents the classical analogues of Fock states and denotes the optical mode 
of the $ m^{th} $ waveguide. It will be shown that Eq. (\ref{eq3}) is equivalent
 to Eq. (\ref{eq2}) in terms of $\Psi_{n}(Z)$.
The set $ \lbrace \vert m \rangle\rbrace $ is called the waveguide number basis and 
$ \Psi_{m}(Z) $ denotes the amplitude of electric field in the $ m^{th} $ 
waveguide which depends on the dimensionless propagation distance.
 $ \widehat{a} $ and $ \widehat{a}^{\dagger} $ are 
peculiar translation operators to the left and right, 
respectively, which are defined by $ \widehat{a} \vert m\rangle = \sqrt{m} \vert m-1\rangle $ and $ \widehat{a}^{\dagger} \vert m\rangle = \sqrt{m+1} \vert m+1\rangle $, 
similar to the bosonic annihilation and creation operators in quantum optics.

In terms of waveguide number basis, equation (\ref{eq3}) is rewritten in 
the following form:
 \begin{eqnarray}
  \label{eq4}
  i\sum_{m} \frac{d\Psi_{m}(Z)}{dZ} \vert m\rangle = -\sum_{m} \sqrt{m(m-1)}\Psi_{m}(Z) \vert m-2\rangle 
  \nonumber \\
- \sum_{m}\sqrt{(m+1)(m+2)} \Psi_{m}(Z) \vert m+2\rangle. \hspace*{1.8cm}
\end{eqnarray}
The orthogonality of waveguide number basis $(\langle m\vert k\rangle =\delta_{m,k})$
 applied to Eq. (\ref{eq4}) leads to,

\begin{eqnarray}
\label{eq5}
i \frac{d\Psi_{k}(Z)}{dZ} + \sqrt{(k+1)(k+2)}\Psi_{k+2}(Z) \hspace*{1.5cm}
\nonumber \\
+\sqrt{k(k-1)} \Psi_{k-2}(Z)=0 ,\hspace*{1.7cm}  
\end{eqnarray}
which justifies the equivalence of Eq. (\ref{eq3}) and (\ref{eq2}). Therefore, it is
sufficient to solve Eq. (\ref{eq3}).

The solution of Eq. (\ref{eq3}) can be written as:
\begin{eqnarray}
\label{eq6}
\Phi (Z) = \exp[iZ(\widehat{a}^{2}+
{\widehat{a}^{\dagger}{}}^{2})]
\Phi (Z=0)= \widehat{S}(-2iZ)\Phi (0),
\hspace*{0.3cm}
\end{eqnarray}
in which $ \widehat{S}(-2iZ) $ is the squeeze operator 
$ \widehat{S}(\xi) = \exp[\frac{1}{2}( \xi^{*}\widehat{a}^{2}-\xi{\widehat{a}^{\dagger}{}}^{2})]$ 
with purely imaginary squeezing parameter, $ \xi = -2iZ $.
The squeezed amplitude is twice as the dimensionless 
propagation length and the squeezed phase is $-\frac{\pi}{2}$.
By employing the disentangling theorem, the squeezed 
operator is given in the following form \cite{Walls, Puri}:
\begin{eqnarray}
\label{eq7}
\widehat{S}(-2iZ)=\frac{1}{\sqrt{\cosh(2Z)}}\exp[\frac{i}{2}(\tanh(2Z)){\widehat{a}^{\dagger}{}}^{2}]
\nonumber \\
\times \exp[-\ln(\cosh(2Z)){\widehat{a}^{\dagger}}\widehat{a}]\exp[\frac{i}{2}(\tanh(2Z))\widehat{a}^{2}].
\end{eqnarray}

If we excite the waveguide array by injecting light beam at the $ n^{th}$
waveguide, the amplitude of light in the $ l^{th} $ 
waveguide at position $ Z $ along the propagation direction would be 
$\Psi_{l}^{(n)}(Z) = \langle l\vert  \widehat{S}(-2iZ)\vert n \rangle $.
% After some straightforward calculations $ \Psi_{l}^{(n)}(Z) $ is written in closed form as:
After some straightforward calculations, for even values of $ \vert n-l\vert $,
$ \Psi_{l}^{(n)}(Z) $ 
can be written in the following closed form:
\begin{eqnarray}
\label{eq8}
\Psi_{l}^{(n)}(Z) =\frac{\sqrt{n!
l!}[\frac{-i}{2}\tanh(2Z)]^{\frac{n+l}{2}}}{\sqrt{\cosh(2Z)}} \hspace*{2.5cm}
\nonumber \\
\hspace*{0.4cm} \times  \sum_{m=0}^{M} \frac{(-2i)^{2m}}{[\sinh(2Z)]^{2m}
[2m]![\frac{l}{2}-m]![\frac{n}{2}-m)]!} , \hspace*{0.5cm}
  \end{eqnarray}
while, $\Psi_{l}^{(n)}(Z) = 0$ if $ \vert n-l\vert$ is equal to an odd number.
Moreover, $M=Int[Min(\frac{n}{2},\frac{l}{2})]$ where 
$Int [x] $ means the integer part of a real number $ x $. 

%%%%%%%%%%%%%%
\begin{figure}[b!]
\includegraphics[angle=0,height=0.8\linewidth, width=0.9\linewidth]{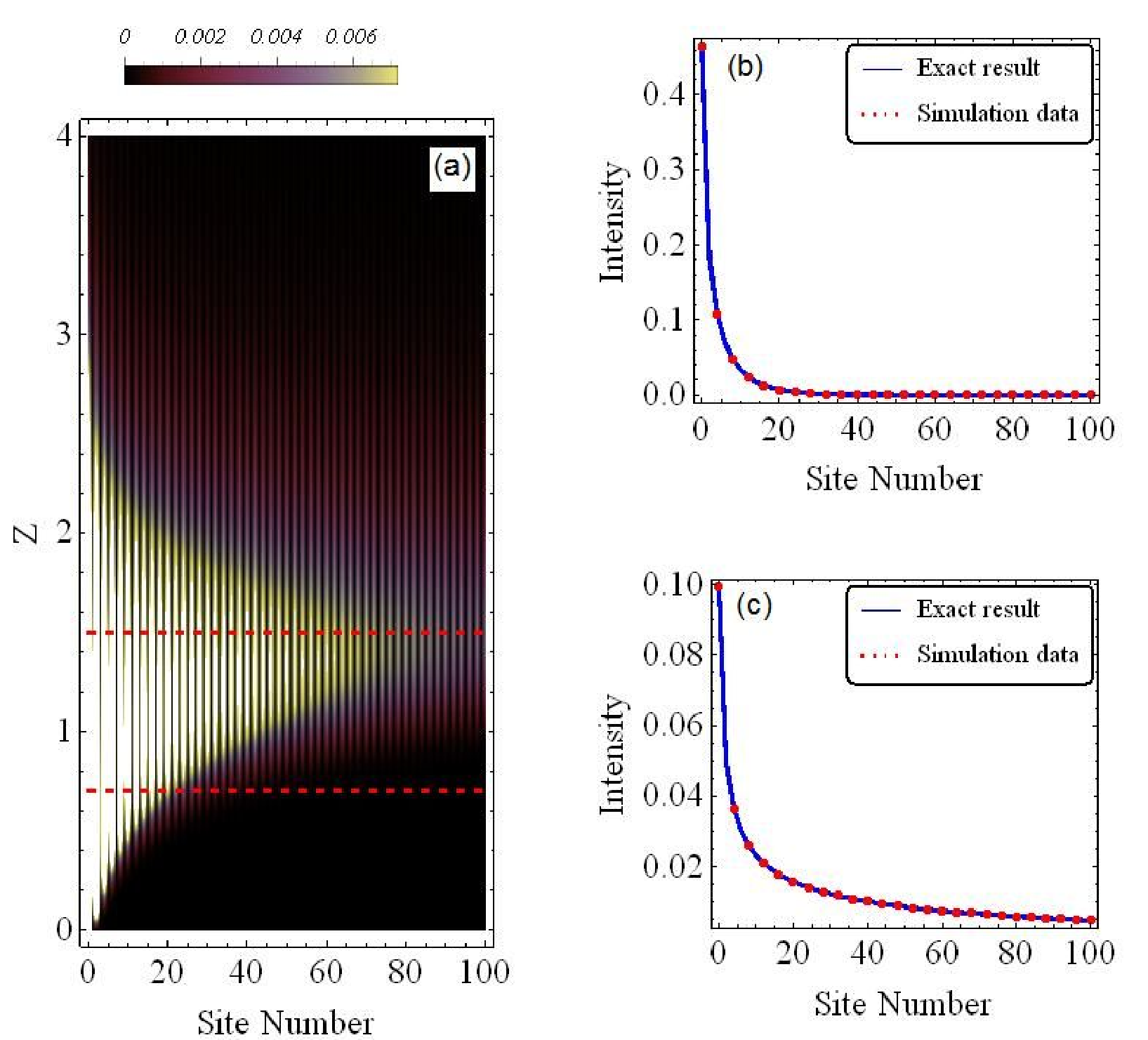}
\caption{(color online)
a: Light intensity distribution for an initial excitation at $n$= 0.
The light intensity profile versus site number at b: $Z$= 0.7 and c: $Z$=
1.5, solid lines are exact results and red circles comes from Runge Kutta
Fehlberg numerical simulation.}
\label{fig2}
\end{figure}
%%%%%%%%%%%%
\begin{figure}[t!]
\includegraphics[angle=0,height=0.9\linewidth, width=1.0\linewidth]{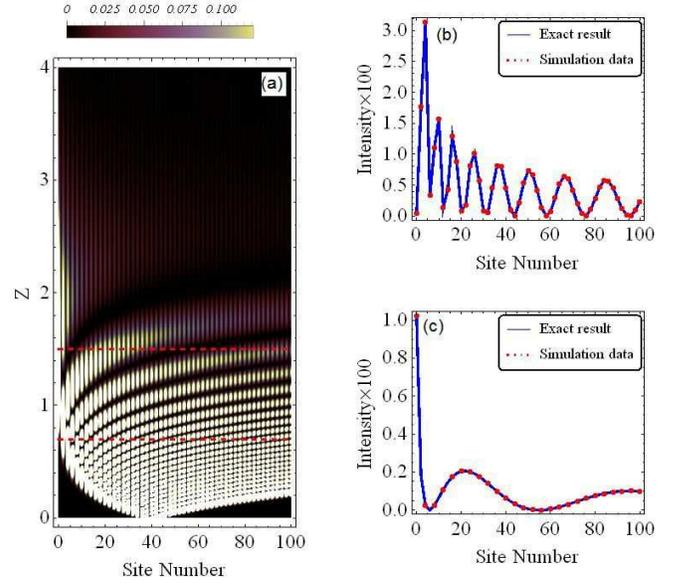}
\caption{(color online)
a: Light intensity distribution for an initial excitation at $n$= 20.
The light intensity profile versus site number at b: $Z$= 0.7 and c: $Z$=
1.5, solid lines are exact results and red circles comes from Runge Kutta
Fehlberg numerical simulation.}
\label{fig3}
\end{figure}
%%%%%%%%%%%%%%%%%
  
The light intensity in the $ l^{th} $ waveguide, at position $ Z $ along the
propagation direction is 
$ I_{l}^{(n)}(Z) = \mid \Psi_{l}^{(n)}(Z) \mid^{2} $.
The light intensity distribution is similar to the photon number
distribution for squeeze number states versus time in quantum optics 
\cite{Squeeze1989, Knight, Walls, Puri}.
If light is injected in the first even labeled waveguide $ (n= 0) $, the light
intensity distribution
is similar to the squeezed vacuum photon distribution. For this
case $ M=0 $, 
and the light intensity distribution is reduced to the following form:
\begin{eqnarray}
%\begin{aligned}
\label{eq9}
I_{l}^{(n=0)}(Z) =\frac{l![\frac{-1}{2}\tanh(2Z)]^{l}}{(\frac{l}{2})!\cosh(2Z)} 
\cos^{2}(\frac{l\pi} {2}).
%\end{aligned}
  \end{eqnarray}
Fig. \ref{fig2}(a) shows the light intensity distribution in lower (even
labeled) array 
which is similar to the photon number distribution of squeezed vacuum state.
Fig. \ref{fig2}(b) and Fig. \ref{fig2}(c) show the intensity versus even labels of
waveguides
at $Z=0.7$ and $Z=1.5$, respectively.
To verify our analytical results, the system of Eq.(\ref{eq2}) for 
10000 guide are solved numerically by Runge Kutta Fehlberg method.
The results of numerical simulation have been shown by red circles in
Fig. \ref{fig2}(b) and Fig. \ref{fig2}(c) which show perfect 
agreement with the exact (analytical) solutions.

Fig. \ref{fig3}(a) shows the light intensity distribution when light is injected
in an intermediate
 waveguide ($ n=20 $) in the lower array, at the entrance plane.
A reflection from the (fixed) left boundary is observed
where the
intensity is returned to the waveguides.
The light intensity profiles are depicted at two different propagation
length $Z=0.7$ and $Z=1.5$ in Fig. \ref{fig3}(b) and Fig. \ref{fig3}(c),
respectively.
These profiles show the oscillation of light intensity in the lower array of
waveguides.
 
According to our study, $ \Psi_{l}^{(n)}(Z) $ is interpreted as the impulse
response for this structures. 
Therefore, for an arbitrary distribution of light intensity injected at 
the entrance
plane,
 the light intensity distribution in each waveguides at the propagation 
distance $Z$, is given by:
\begin{eqnarray}
\label{eq10}
\Psi_{l}(Z) = \sum_{n=0}^{\infty} \Psi_{l}^{(n)}(Z) \Psi_{n}(Z=0). \hspace*{0.5cm}
\end{eqnarray} 
   
If the intensity profile at entrance plane is chosen to be a Poisson
distribution such as 
$\Psi_{n}(Z=0)= \frac{\beta^{n}}{\sqrt{n!}}\exp(-\frac{\vert \beta\vert^{2}}{2})$, 
where $\beta=\vert\beta\vert \exp (i \theta)$, 
the light intensity distribution at distance $Z$, 
can be written as follows (for more details, see the appendix):
\begin{eqnarray}
\label{eq11}
I_{n}(Z) = \vert\langle n\vert \widehat{S}(-2iZ)\vert \beta\rangle\vert^{2}= \vert\langle n\vert Z,\beta \rangle\vert^{2} \,\,\,
\nonumber \\
= \frac{[\frac{1}{2}\tanh(2Z)]^{n}}{n!\cosh(2Z)}\exp[-\vert\beta\vert^{2}(1+\sin(2 \theta)\tanh(2Z))]
\nonumber \\
\times \vert H_n[\frac{\beta}{\sqrt{-i\sinh(4Z)}}]\vert^{2}, \hspace*{2.2cm} 
\end{eqnarray}
where $ H_n(x) $ is the Hermite polynomial of order $n$. 
The intensity profile of Eq.(\ref{eq11}) is reminiscent of the squeezed
coherent photon distributions. 

\begin{figure}[t!]
\includegraphics[angle=0,height=0.9\linewidth, width=1.0\linewidth]{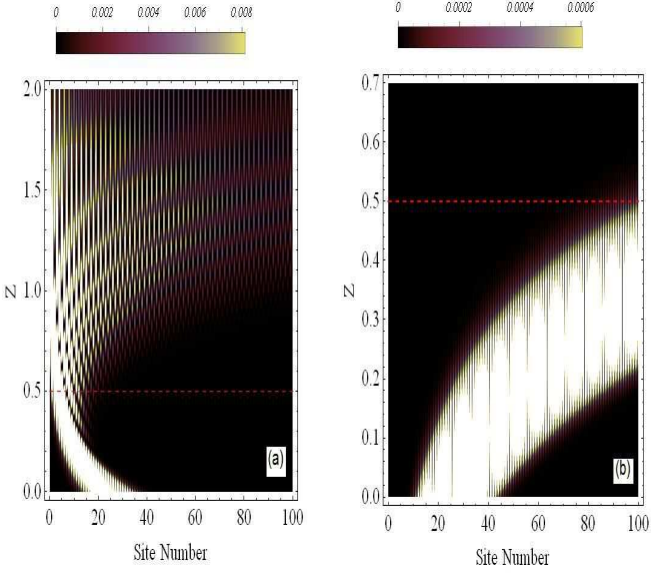}
\caption{(color online)
Light intensity distribution in waveguide array along the waveguide length, 
for $N$= 100, $\vert\beta\vert$= 4.0,  a: $\theta=-\frac{\pi}{4}$,
b: $\theta=+\frac{\pi}{4}$.}
\label{fig4}
\end{figure}

Fig. \ref{fig4} shows the light intensity distribution along Z and versus
site number, 
if a coherent light distribution is injected at the initial plane (Z=0). 
Fig. \ref{fig4}(a) and Fig. \ref{fig4}(b) present the light intensity 
distribution for two different phase of the initial light,
namely: $\theta=-\frac{\pi}{4}$ and $\theta=+\frac{\pi}{4}$, respectively.
These figures show that the reflection from the left boundary of the
semi-infinite array,
can be occurred only for the negative initial phase.

The authors of Refs. \cite{Fock2010, FockPRL} proposed waveguide
lattices which provide the light intensity distribution
for a coherent state by injecting an initial beam in the first
guide of lattice at the entrance plane. 
However, the interesting point of our work is to 
design a self-consistent structure implementing two different arrays, 
to provide the squeezed coherent light intensity distribution. 
The squeezed coherent light distribution is established when 
the initial beam is exposed only to one of the waveguides at the entrance plane.
This can be applied to simulate classically
some phenomena related to the squeezed coherent states in quantum optics.
 
As mentioned before, if the exposing beam is applied on an even labeled
waveguide, light propagates in the lower array, while an injection of light on
the odd labeled waveguide the propagation is on the upper array.
The coherent state is
a superposition of both even and odd states. For a coherent state, 
light propagates
both in the upper and lower arrays. The intensity distribution would be similar
to the photon distribution
of coherent squeezed state \cite{Squeeze1989, Knight, Walls, Puri}.

\section{Spatial Bloch oscillation \label{sbo}}

In order to study the spatial Bloch oscillation in waveguide arrays,
a linear transverse gradient is added to the  propagation constants i.e. propagation 
constants of waveguides are given by $ K_{n} = K_{0}+\Delta n $ where $\Delta$ is a constant.
In the presence of the additional term ($\Delta n$), Eq. (\ref{eq2})
is written in the following form:
\begin{eqnarray}
\label{eq12}
i\frac{d \Psi_{\alpha, n}}{dZ}(Z)+\alpha n\Psi_{\alpha, n}(Z)+\sqrt{n(n-1)}
\Psi_{\alpha, n-2}(Z) \hspace*{0.5cm} 	
\nonumber \\
+\sqrt{(n+1)(n+2)}\Psi_{\alpha, n+2}(Z)=0,\hspace*{0.4cm}  n =0,1,2,...\hspace*{0.5cm} 	      
\end{eqnarray}
where $ \alpha=\frac{\Delta}{C_{1}} $. 
As shown for Eq. (\ref{eq2}), the following equation is equivalent to
Eq. (\ref{eq12})
\begin{eqnarray}
\label{eq13}
i\frac{d \Phi}{dZ}(Z) = -(\widehat{a}^{2}+\alpha \widehat{a}^{\dagger}\widehat{a}
+{\widehat{a}^{\dagger}{}}^2)\Phi (Z),
\end{eqnarray}
 if $ \Phi(Z) $ is expanded
in the orthonormal waveguide number basis. 
%A straightforward calculation shows that the following form of $\Phi (Z)$ 
%gives the solution of Eq. (\ref{eq13})
The solution of Eq. (\ref{eq13}) is:
\begin{eqnarray}
\label{eq14}
\Phi (Z) = \exp[iZ(\widehat{a}^{2}+{\widehat{a}^{\dagger}{}}^{2}
+\alpha\widehat{a}^{\dagger}\widehat{a})]\Phi (Z=0)
\nonumber \\
= \widehat{S_{\alpha}}(Z)\Phi (0), \hspace*{0.5cm}
\end{eqnarray}
where $ \widehat{S}_{\alpha}(Z) $ can be expressed in terms of
SU(1, 1) Lie generators. The generators are defined by
$ \widehat{K_{+}}=\frac{{\widehat{a}^{\dagger}{}}^{2}}{2} $ , 
$ \widehat{K_{-}}=\frac{\widehat{a}^{2}}{2} $ 
and $ \widehat{K_{z}}=\frac{1}{2} (\widehat{a}^{\dagger}\widehat{a}+\frac{1}{2}) $
which satisfy the following algebra:
\begin{eqnarray}
  \label{eq15}
[\widehat{K_{+}}, \widehat{K_{-}}] = -2\widehat{K_{z}} ;\hspace*{0.5cm}
 [\widehat{K_{z}}, \widehat{K_{\pm}}] = \pm\widehat{K_{\pm}}. 
\end{eqnarray}
According to the structure of Lie algebra and employing the disentangling theorem 
we write $ \widehat{S}_{\alpha}(Z) $ as the product 
of three exponential forms  \cite{Puri}:
\begin{eqnarray}
\label{eq16}
\widehat{S_{\alpha}}(Z) = 
\exp[-\frac{i\alpha Z}{2}]
\exp[iZ(2\widehat{K_{+}}
+2\widehat{K_{-}}+2\alpha 
\widehat{K_{z}})] \hspace*{0.4cm}
\nonumber\\
= \exp[-\frac{i\alpha Z}{2}] 
\exp[\phi(Z)\widehat{K_{+}}]
\exp[\phi_{z}(Z)\widehat{K_{z}}]
\exp[\phi (Z)\widehat{K_{-}}] ,
\nonumber\\
\end{eqnarray}
where
\begin{eqnarray}
\label{eq17}
\phi(Z)=\frac{-2\sinh(\Gamma Z)}{i\Gamma \cosh(\Gamma Z)+\alpha \sinh(\Gamma Z)}\hspace*{2.5cm}
\nonumber \\
\phi_{z}(Z)=-2\ln[\cosh(\Gamma Z)+\frac{\alpha}{i\Gamma}\sinh(\Gamma Z)]. \hspace*{1.5cm}
\end{eqnarray}

Here $ \Gamma =\sqrt{4-\alpha^{2}} $ which is real for
 $ \alpha <2 $ while it is purely imaginary for $\alpha >2 $.

The light intensity in the $ l^{th} $ waveguide at position $Z$, when
the array is excited by an input at the $ n^{th} $ waveguide is given by
$I_{l}^{(n)}= 
\vert\langle l \vert \widehat{S}_{\alpha}(Z) \vert n\rangle\vert^{2}$,
and can be obtained by employing the normal form factorized evolution operator 
(equation (\ref{eq16})). 
More insight on the product form of $\widehat{S}_{\alpha}(Z)$ reveals that
even (odd) waveguides are coupled to even (odd) ones. 
Because, the SU(1, 1) algebra confirms that the operation of $\widehat{K_{+}}$ 
or  $\widehat{K_{-}}$ on an even (odd) state leads to another even (odd) state.
Hence, the amplitude of light in the $ l^{th} $ waveguide  
$ (\Psi_{\alpha, l}^{(n)}(Z) )$ is obtained to be
\begin{eqnarray}
\label{eq18}
\Psi_{\alpha, l}^{(n)}(Z) = \langle l\vert  \widehat{S}_{\alpha}(Z)\vert n\rangle \hspace*{4.5cm}
\nonumber \\
=\frac{\sqrt{i\Gamma(n!) (l!)}[-\sinh(\Gamma Z)]^{\frac{n+l}{2}}
e^{-\frac{i\alpha Z}{2}}}{[\alpha \sinh(\Gamma Z)+
i\Gamma \cosh(\Gamma Z)]^{\frac{n+l+1}{2}}}\hspace*{1.1cm}
\nonumber \\
\times  \sum_{m=0}^{M} \frac{(-i\Gamma)^{2m}}
{[\sinh(\Gamma Z)]^{2m}[2m]![\frac{l}{2}-m]![\frac{n}{2}-m)]!},\hspace*{0.4cm}
\end{eqnarray}
for $ \vert n-l\vert = \mbox{even} $ and $ \Psi_{l}^{(n)}(Z) = 0 $ 
whenever $ \vert n-l\vert = \mbox{odd}$. It is straightforward to show 
that the solution presented in Eq. (\ref{eq18})
is reduced to Eq. (\ref{eq9}) for $ \alpha=0 $.

 \begin{figure}[b!]
\includegraphics[angle=0,height=0.9\linewidth, width=1.0\linewidth]{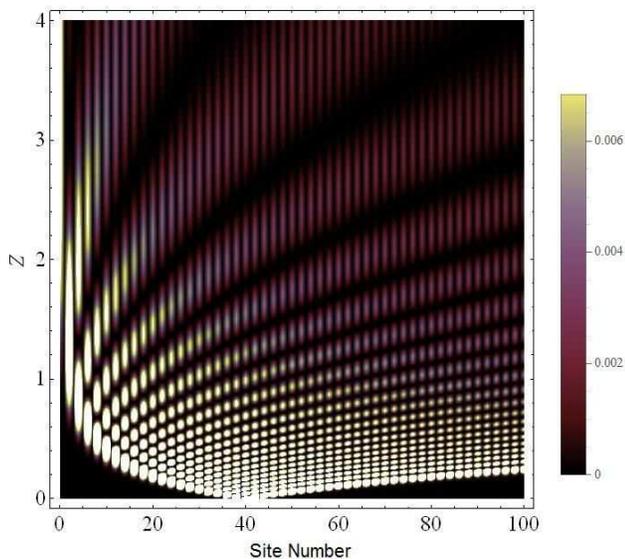}
\caption{(color online)
The profile of light intensity at the critical point $\alpha$= 2.0 where
the initial beam is exposed at $n=40$ for the semi-infinite array of waveguides.
}
\label{alpha2}
\end{figure} 
  
For $\alpha<2$, $\Gamma$ is a real parameter, hence, the profile of light
intensity is similar to what is presented already in Fig. \ref{fig3}
in the absence of linear transverse gradient in the propagation constants.
If we concentrate our attention to the light propagation along Z direction
within a single waveguide, its long distance behavior is decaying
exponentially, i.e. $I_{\alpha<2, l}^{(n)}(Z) \sim exp(-\Gamma Z)$ for
$Z\gg1$. 
It is reminiscent the overdamped behavior of the underlying system.
However, for $\alpha>2$, $\Gamma$ is a purely imaginary parameter which will
change the behavior of our system (as will be discussed). Therefore, we 
anticipate a phase transition at the critical parameter $\alpha_c=2$.

At $\alpha=\alpha_c=2$, the parameter $\Gamma$ is zero which necessitates to 
evaluate Eq.(\ref{eq18}) in the limit $\Gamma\rightarrow 0$, we get
\begin{eqnarray}
\label{eq19}
\Psi_{\alpha=2, l}^{(n)}(Z) =\frac{(iZ)^{\frac{n+l}{2}}\sqrt{n!
l!}e^{-iZ}}{[1-2iZ]^{\frac{n+l+1}{2}}} \hspace*{2.0cm} 
\nonumber \\
\times \sum_{m=0}^{M} \frac{1}{[
iZ]^{2m}[2m]![\frac{l}{2}-m]![\frac{n}{2}-m)]!}, \hspace*{0.7cm}
\end{eqnarray}
for $\vert n-l\vert = \mbox{even}$ and $\Psi_{l}^{(n)}(Z) = 0$ 
for $\vert n-l\vert =\mbox{odd}$. 
The intensity profile at the critical point
($\alpha=\alpha_c$) is
plotted in Fig. \ref{alpha2}. The long distance behavior ($Z\gg1$) of
intensity profile for any waveguide decays algebraically, 
i.e. $I_{\alpha_c, l}^{(n)}(Z) \sim \frac{1}{Z}$. This is the typical 
behavior at a critical point where fluctuations of all scales contribute
to the phenomenon. Here, the correlations decay algebraically which states
that the diffraction exists on all length scales.

\begin{figure}[b!]
\includegraphics[angle=0,height=0.9\linewidth, width=1.0\linewidth]{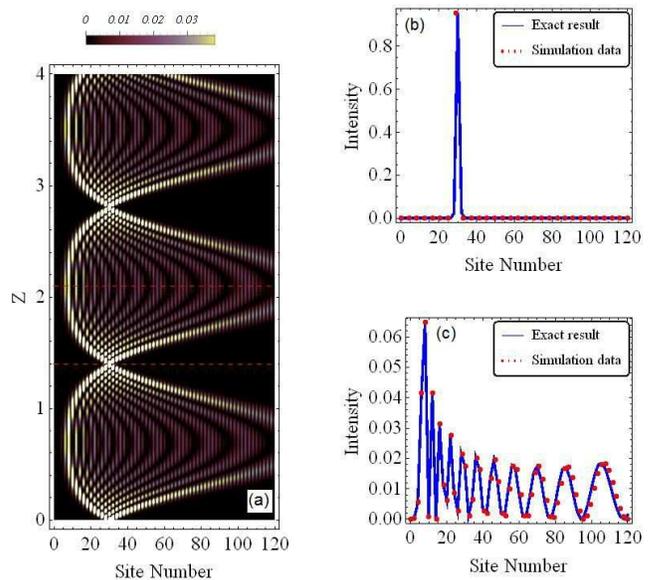}
\caption{(color online)
a: Light intensity distribution of Bloch-oscillation in the 
semi-infinite array of waveguides
when the initial beam is exposed at $n$= 30 for $\alpha$= 3.0.
The intensity profile versus site number at b: Z=1.4 and c: Z=2.1,
solid lines are exact analytical results and red circles are the results of
Runge Kutta
Fehlberg numerical simulation.
}
\label{fig6}
\end{figure} 

Nevertheless, for $ \alpha > 2 $, $ \Gamma $ is purely imaginary and the
hyperbolic functions in Eq. (\ref{eq18}) are converted to the periodic
trigonometric functions where the spatial Bloch oscillation comes up. 
The spatial frequency of this oscillation is 
$Z_{T}=Cz_{T}=\frac{\pi}{\vert\Gamma\vert} = \frac{\pi}{\sqrt{\alpha^{2}-4}}$. 
 Fig. \ref{fig6} shows the intensity pattern of this periodic propagation
for $ \alpha=3.0 $ for which the spatial frequency is $ Z_{T}\cong 1.405 $. 

The spatial Bloch oscillation turns out from the interplay between discrete
diffraction,
Bragg diffraction and surface reflection effects. In our model, due to the
increase
of the coupling coefficients by increasing the site numbers, discrete
diffraction causes
the expansion of light over large numbers of waveguides during propagation 
along waveguides. 
The role of Bragg diffraction appears when the phase
differences between
neighboring waveguides are equal to multiple of $ \pi $. 
This condition can be satisfied at 
certain propagation lengths. At these points the expansion of
light is terminated and light returns to the waveguides with lower 
propagation constants.
Eq. (\ref{eq12}) shows that for high site labels $(n\gg 1) $, both the coupling
coefficients and propagation 
constants are roughly proportional to the site index $ n $.
Therefore, competition of discrete diffraction and Bragg
diffraction causes the existence 
of a critical value for $ \alpha $. For $ \alpha $ less than its critical value
($ \alpha\leq \alpha_{c} $) the Bragg diffraction is
suppressed by discrete
diffraction,
while for higher $ \alpha $  ($ > \alpha_{c}$) the Bragg
diffraction
is dominated.
Moreover, the surface reflection causes repulsion
at $ n = 0 $ boundary. For $ \alpha > \alpha_{c} $, Bloch oscillation occurs due
to the reflection at high ($n\gg1$) and $ n = 0 $ waveguides. Such a critical
value does not 
exists in Fock-Gluaber lattices where the coupling coefficients are 
approximately proportional to the square root of the site label ($\sqrt{n}$),
while
the propagation constants are proportional to the waveguide label ($n$).
Hence, 
independent of how much the value of $ \alpha $ is, the Bragg reflection is
dominated 
and Bloch oscillation occurs.

\begin{figure}[t!]
\includegraphics[angle=0,height=0.9\linewidth, width=1.0\linewidth]{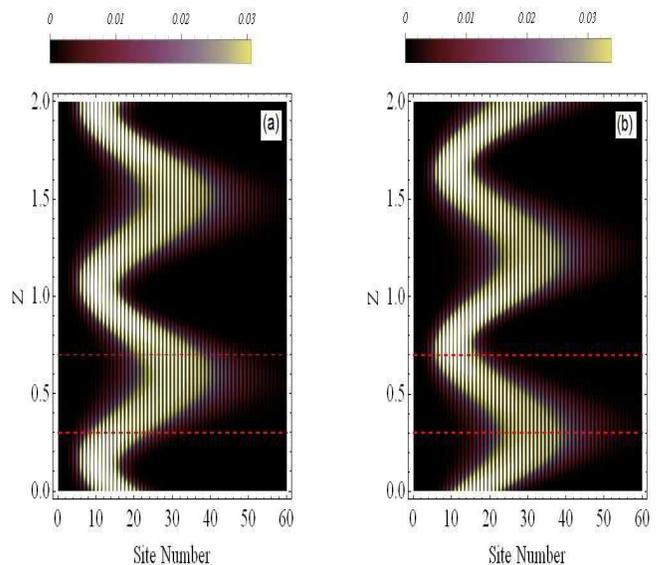}
\caption{(color online)
Light intensity distribution in waveguide array along the waveguide length, for
$N$= 60,
$\alpha$= 4.0, $\vert\beta\vert$= 4.0,  a: $\theta=-\frac{\pi}{4}$,  b:
$\theta=+\frac{\pi}{4}$.
}
\label{fig7}
\end{figure}

$ \Psi_{\alpha, l}^{(n)}(Z) $ is the impulse response for such structures.
For an arbitrary distribution of light intensity injected at $Z=0$, the
light
intensity distribution in each waveguides at propagation distance $ Z $,
can be calculated as follows:
\begin{eqnarray}
\label{eq20}
\Psi_{\alpha, l}(Z) = \sum_{n=0}^{\infty} \Psi_{\alpha, l}^{(n)}(Z) \Psi_{n}(Z=0) . \,\,\,\,\,\,\,\,
\end{eqnarray}
If the light intensity distribution at the entrance plane is chosen from a
Poisson distribution, i.e. 
$ \Psi_{n}(Z=0)=e^{-\frac{\vert \beta \vert^{2}}{2}}\frac{\beta^{n}}{\sqrt{n!}} $, 
where $ \beta=\vert\beta\vert e^ {i \theta} $, the light intensity distribution at
propagation distance $ Z $, 
can be written as follows 
(more details are presented in the appendix):
\begin{eqnarray}
\label{eq21}
I_{n}(Z) =\vert\langle n\vert  \widehat{S_{\alpha}}(Z)
\vert \beta \rangle\vert^{2} = \vert\langle n\vert Z_{\alpha},
\beta \rangle\vert^{2} \hspace*{1.3cm}
\nonumber \\
= \frac{\vert\frac{\nu^{'}}{2\mu^{'}}\vert^{n}}{n!\vert \mu^{'}
\vert}\exp[-\vert\beta\vert^{2}(1+Re(\frac{\nu^{'}e^{2i\theta}}{\mu^{'}})]\hspace*{0.3cm}
\nonumber \\
\times \vert H_n[\frac{\beta}{\sqrt{2\mu^{'}\nu^{'}}}]\vert^{2}. \hspace*{0.8cm}
\end{eqnarray}
Here $\nu^{'} = \frac{2\nu}{\Gamma}$ , $\mu^{'} =\mu+
\frac{\alpha\nu}{\Gamma}$, 
$\nu = -i\sinh(\Gamma Z)$ and $\mu = \cosh(\Gamma Z)$. We call 
$\vert Z_{\alpha},\beta \rangle $ as the generalized coherent squeezed states. 
Fig. \ref{fig7} shows the light intensity distribution for
$ \alpha > \alpha_{c} $, if the coherent light intensity distribution is 
injected at $Z=0$ plane. The spatial-Bloch oscillation is seen easily in
this figure.
Fig.\ref{fig7}(a) and Fig.\ref{fig7}(b) show the dependence of light intensity 
distribution on the initial phase of light.

\section{Summary \label{summary}}

We proposed the classical analogues of quantum
\textit{squeezed number} and \textit{squeezed coherent states}  in a semi-infinite
lattice of waveguides with an appropriate tuning of the coupling coefficients.
We have obtained the closed analytic form for the 
light intensity of any waveguide at position $Z$ along its length, 
regardless of which waveguide has been exposed by an initial beam.
The result has been extended to get the intensity profile for an
arbitrary intensity distribution at the entrance of waveguides.
Adding a linear gradient ($\alpha$) to the propagation constant of 
the $n$-th waveguide
leads to a phase transition between two different behaviors. For $\alpha <2$,
we observed a pattern which comes from discrete diffraction similar to the
$\alpha=0$ case 
(Fig. \ref{fig3}) while
for $\alpha >2$ we observe the  \textit{spatial Bloch oscillation } 
in the array of waveguides (Fig. \ref{fig6}).

The nature of phase transition is related to the interplay between discrete 
diffraction, which tends to expand the light over large numbers of waveguides, 
and Bragg reflection, which causes the light intensity to return to waveguides
with lower propagation constants. 
Meanwhile, the reflection at $ n=0 $ boundary causes the light to return to the
waveguides with increasing labels. Hence, an oscillation appears. 
We have also found the close form of light intensity distribution,
in the presence of linear gradient of refraction index, 
when a Poisson light distribution is chosen for the initial beam which is
a classical simulation of generalized squeezed coherent states.

We propose that these fully integrable lattices provide new opportunities to
study some interesting phenomena in quantum optics and
condensed matter physics, such as \textit{photon correlations} and
\textit{quantum phase transitions}, respectively. We have obtained
the long distance ($Z\gg1$) intensity profile which falls off exponentially
for $\alpha <2$ and an algebraic decay at the critical point  $\alpha=2$.
This is similar to the spatial behavior of correlation functions in a magnetic
system close to quantum critical point. However, the observation and measurement of
light intensity in an array of waveguides are much simpler than the 
corresponding counterpart in a magnetic system.

%\section*{Acknowledgment}

\appendix*
%\section*{Appendix}
\section{}
\label{App}

In order to obtain Eqs. (\ref{eq11}) and (\ref{eq21}), 
we start with more general case when $ \alpha\neq 0 $. 
For $ \alpha =0 $, Eq. (\ref{eq21}) can be converted to
Eq. (\ref{eq11}), so it is sufficient to get Eq. (\ref{eq21}). 
 
We define $ \widehat{a}\vert 0\rangle =0 $ which leads to
\begin{equation}
\widehat{S_{\alpha}}(Z)\widehat{D}(\beta) \widehat{a}\widehat{D}^{\dagger} (\beta)\widehat{S_{\alpha}}^{\dagger}(Z) \vert Z_{\alpha},\beta \rangle = 0.
\end{equation}
Here we have defined $\vert Z_{\alpha},\beta\rangle =\widehat{S_{\alpha}}(Z)\widehat{D}(\beta)\vert 0\rangle $.
The implementation of $ \widehat{D}(\beta) \widehat{a} \widehat{D}^{\dagger}(\beta) = \widehat{a}-\beta $ leads to
$\widehat{S_{\alpha}}(Z) \widehat{a} \widehat{S_{\alpha}}^{\dagger}(Z)
\vert Z_{\alpha},\beta \rangle =\beta \vert Z_{\alpha},\beta \rangle$.

After some lengthy but straightforward calculations we obtain:
$ \widehat{S_{\alpha}}(Z)\widehat{a}\widehat{S_{\alpha}}^{\dagger}(Z)=\mu^{'}\widehat{a}+\nu^{'}\widehat{a}^{\dagger} $.
So,
\begin{eqnarray}
\mu^{'}\widehat{a}+\nu^{'}\widehat{a}^{\dagger}\vert Z_{\alpha},
\beta \rangle =\beta\vert Z_{\alpha},\beta\rangle. \,\,\,\,\,\,\,\,\label{A-1}
\end{eqnarray}
We expand the generalized coherent squeezed states in the Fock bases 
$\vert Z_{\alpha},\beta\rangle = \sum_{n} C_{n}\vert n \rangle $. 
In this bases we arrive at the following relation
\begin{eqnarray}
\sum_{n}[\mu^{'}C_{n}\sqrt{n}\vert n-1 \rangle +\nu^{'}C_{n}\sqrt{n+1}\vert n+1 \rangle ]
\nonumber \\  
= \gamma \sum_{n} C_{n}\vert n \rangle ,\label{A-2}
\end{eqnarray}
We define
$C_{n}\equiv \frac{N}{\sqrt{\mu^{'}}}[\frac{\nu^{'}}{2\mu^{'}}]^{\frac{n}{2}}f_{n}(x)$, 
and replace it in Eq. (\ref{A-2}) which gives
\begin{eqnarray}
\sqrt{n+1}f_{n+1}+2\sqrt{n}f_{n-1}-\frac{2\beta}{\sqrt{2\mu^{'}\nu^{'}}}f_{n}=0, \label{A-3}  
\end{eqnarray}

Eq. (\ref{A-3}) is similar to the recursion relation of Hermite polynomials if we 
consider $ f_{n} = \frac{1}{\sqrt{n!}} H_{n}(x) $ and 
$ x = \frac{\beta}{\sqrt{2\mu^{'}\nu^{'}}} $. 
Thus, the expansion coefficients $ C_{n} $ can be written as   
\begin{eqnarray} 
 C_{n}=\frac{N}{\sqrt{n! \mu^{'}}}[\frac{\nu^{'}}{2\mu^{'}}]^{\frac{n}{2}}H_{n}(\frac{\beta}{\sqrt{2\mu^{'}\nu^{'}}}), 
 \label{A-4}
 \end{eqnarray} 
which gives $ C_0 = \frac{N}{\sqrt{\mu^{'}}} $. On the other hand 
$C_0 = \langle 0 \vert Z_{\alpha}, \beta\rangle  $ which leads to 
$ N = \sqrt{\mu^{'}} \langle 0\vert Z_{\alpha},\beta\rangle $.
Moreover, we have
\begin{eqnarray} 
\langle 0 \vert Z_{\alpha},\beta\rangle = \frac{e^{-\frac{i\alpha Z}{2}}}{\sqrt{\mu^{'}}}
\exp[-\frac{\vert \beta \vert ^{2}}{2}-\frac{1}{2}{\beta}^{2}(\frac{\nu^{'}}{\mu^{'}}) ]. \label{A-5}
\end{eqnarray}

Therefore, by using Eqs. (\ref{A-4}) and (\ref{A-5}) for  $ C_{n} $, 
it is straightforward to reach Eq. (\ref{eq21}).


\begin{references}

%1
\bibitem{Repkiv}
I. L. Garanovich, S. Longhi, A. A. Sukhorukov, Y. S. Kivshar,
Physics Reports {\bf 518}, 1 (2012). 
%Ivan L. Garanovich, Stefano Longhi, Andrey A. Sukhorukov, Yuri S. Kivshar
%% "Light propagation and localization in modulated photonic lattices and waveguides",
% Physics Reports {\bf 518} (2012) 1-79. 

%2 
\bibitem{Replon}
S. Longhi, Laser and Photon. Rev. {\bf 3}, 3 (2009).
%Laser and Photon. Rev. {\bf 3}, 3, (2009) 243–261.
%% "Quantum-optical analogies using photonic structures",

%3
\bibitem{RepLed}
F. Lederer, G. I. Stegeman, D. N. Christodoulides, G. Assanto, M. Segev, Y. Silberberg,
 Physics Reports {\bf 463}, 1 (2008).
%Falk Lederer, George I. Stegeman, Demetri N. Christodoulides, Gaetano Assanto, Moti Segev, Yaron Silberberg
% Physics Reports {\bf 463} (2008) 1-126.
  %% "Discrete solitons in optics",
  
%4
\bibitem{Nat2003}
D. N. Christodoulides, F. Lederer and Y. Silberberg, Nature {\bf 244}, 817 (2003).
%Demetrios N. Christodoulides, Falk Lederer & Yaron Silberberg
%Nature {\bf 244} (2003) 817-823.
  %% "Discretizing light behaviour in linear and nonlinear waveguide lattices",
  
%5
\bibitem{DL1986}
D. H. Dunlap and V. M. Kenkre
%D. H. Dunlap and V. M. Kenkre
%% "Dynamic localization of a charged particle moving under the influence of an electric field",
  Phys. Rev. B {\bf 34}, 3625 (1986).

%6
\bibitem{Nat2009}
A. Szameit {\em et al.}, Nature {\bf 5}, 271 (2009).
%Alexander Szameit, Ivan L. Garanovich, Matthias Heinrich, Andrey A. Sukhorukov, Felix Dreisow, Thomas Pertsch, Stefan Nolte, Andreas Tünnermann and Yuri S. Kivshar
 %% "Polychromatic dynamic localization in curved photonic lattices",
%Nature {\bf 5} (2009) 271-275.

%7
\bibitem{Nat2007}
 T. Schwartz, G. Bartal, S. Fishman, and M. Segev, Nature {\bf 466}, 52 (2007).
%% "Transport and Anderson localization in disordered two-dimensional photonic lattices",
%Nature {\bf 466} (2007) 52-55.

%8
 \bibitem{PRL2008}
 Y. Lahini {\em et al.},
%  Y. Lahini, A. Avidan, F. Pozzi, M. Sorel, R. Morandotti, D. N. Christodoulides, and Y. Silberberg,
%% "Anderson Localization and Nonlinearity in One-Dimensional Disordered Photonic Lattices",
Phys. Rev. Lett. {\bf 100}, 013906 (2008).


%9
\bibitem{Fock2010}
A. Perez-Leija, H. Moya-Cessa, A. Szameit and D. N. Christodoulides,
%% Armando Perez-Leija, Hector Moya-Cessa, Alexander Szameit, and Demetrios N. Christodoulides
%%  " Glauber–Fock photonic lattices",
Opt. Lett. {\bf 35}, 14 (2010).
 
 
% 10
 \bibitem{FockPRL}
R. Keil {\em et al.}
%Robert Keil, Armando Perez-Leija, Felix Dreisow, Matthias Heinrich, Hector Moya-Cessa, Stefan Nolte, Demetrios N. Christodoulides, and Alexander Szameit
%% "Classical Analogue of Displaced Fock States and Quantum Correlations in Glauber-Fock Photonic Lattices",
Phys. Rev. Lett. {\bf 107}, 103601 (2011).
 

%11 
\bibitem{Fock2012}
R. Keil {\em et al.}
%% Robert Keil, Armando Perez-Leija, Parinaz Aleahmad, Hector Moya-Cessa, Stefan Nolte, Demetrios N. Christodoulides, and Alexander Szameit
%%  " Observation of Bloch-like revivals in semi-infinite Glauber–Fock photonic lattices",
 Opt. Lett. {\bf 37}, 18 (2012).
 
 
% 12
 \bibitem{FockPRA}
A. Perez-Leija, R. Keil, A. Szameit, A. F. Abouraddy, 
H. Moya-Cessa and D. N. Christodoulides
Phys. Rev. A {\bf 85}, 013848 (2012).
%Armando Perez-Leija, Robert Keil, Alexander Szameit, Ayman F. Abouraddy, Hector Moya-Cessa, and Demetrios N. Christodoulides
%% "Tailoring the correlation and anticorrelation behavior of path-entangled photons in Glauber-Fock oscillator lattices",
  

%13
 \bibitem{Bloch1928}
F. Bloch, Z. Phys. {\bf 52}, 555 (1928).


 
% 14
 \bibitem{Blockcurved}
G. Lenz, I. Talanina and C. M. de Sterke,
%G. Lenz, I. Talanina and C. Martijn de Sterke
%% "Bloch Oscillations in an Array of Curved Optical Waveguides",
Phys. Rev. Lett. {\bf 83}, 963 (1999).
 
 
% 15
  \bibitem{BlockTemp}
T. Pertsch, P. Dannberg, W. Elflein, and A. Brauer, F. Lederer,
%T. Pertsch, P. Dannberg, W. Elflein, and A. Bräuer, F. Lederer
%% "Optical Bloch Oscillations in Temperature Tuned Waveguide Arrays",
Phys. Rev. Lett. {\bf 83}, 4752 (1999).
 
%16 
\bibitem{Nonlin bloch}
R. Morandotti, U. Peschel, and J. S. Aitchison, H. S. Eisenberg and Y. Silberberg,
%R. Morandotti, U. Peschel, and J. S. Aitchison, H. S. Eisenberg and Y. Silberberg
%% "Experimental Observation of Linear and Nonlinear Optical Bloch Oscillations",
  Phys. Rev. Lett. {\bf 83}, 4756 (1999).
 
 
 %17
\bibitem{Blochcorr.}
F. Dominguez-Adame, V. A. Malyshev, F. A. B. F. de Moura and M. L. Lyra,
% F. Domı´nguez-Adame and V. A. Malyshev, F. A. B. F. de Moura and M. L. Lyra
%% "Bloch-Like Oscillations in a One-Dimensional Lattice with Long-Range Correlated Disorder",
Phys. Rev. Lett. {\bf 91}, 197402 (2003).
 

%18
 \bibitem{Bloch2d2006}
H. Trompeter {\em et al.},
% Henrike Trompeter, Wieslaw Krolikowski, Dragomir N. Neshev, Anton S. Desyatnikov, Andrey A. Sukhorukov, Yuri S. Kivshar, Thomas Pertsch, Ulf Peschel, and Falk Lederer
%% "Bloch Oscillations and Zener Tunneling in Two-Dimensional Photonic Lattices",
Phys. Rev. Lett. {\bf 96}, 053903 (2006).
 
 
 % 19
  \bibitem{Squeeze1989}
M. S. Kim, F. A. M. de Oliveira, and P. L. Knight,
%M. S. Kim, F. A. M. de Oliveira, and P. L. Knight
%% "Properties of squeezed number states and squeezed thermal states",
  Phys. Rev. A {\bf 40}, 2494 (1989).

% 20 
\bibitem{Knight}
C. Gerry, P. Knight,
%Christopher Gerry, Peter Knight,
{\em Introductory Quantum Optics}
 (Cambridge University Press, 2005).
 
 
 %21
\bibitem{Puri}
 R. R. Puri,
{\em Mathematical methods of quantum optics}
(Springer, 2001).
 
 
 
%22
\bibitem{Walls}
D. F. Walls,  G. J. Milburn,
  {\em Quantum Optics}
  (2nd Edition, Springer-Verlag Berlin Heidelberg, 1994).
 

%23
\bibitem{Yariv}
A. Yarive,
{\em Quantum Electronics}
( third ed., Wiley, New York, 1989).


% 23
% \bibitem{JCLong}
%S. Longhi
%%% Stefano Longhi
%%%  " Jaynes –Cummings photonic superlattices",
% Opt. Lett. {\bf 36}, 17 (2011) 3407.
% 

%24
%\bibitem{Zener2006}
%H. Trompeter, T. Pertsch, F. Lederer, D. Michaelis, U. Streppel, and A. Bra¨uer and U. Peschel
%%Henrike Trompeter, Thomas Pertsch, and Falk Lederer, Dirk Michaelis, Ulrich Streppel, and Andreas Bra¨uer, Ulf Peschel
%%% "Visual Observation of Zener Tunneling",
%  Phys. Rev. Lett. {\bf 96}, 023901(2006).
%

%25
%\bibitem{DL2006}
%S. Longhi, M. Marangoni, M. Lobino, R. Ramponi, and P. Laporta, E. Cianci and V. Foglietti
%%S. Longhi, M. Marangoni, M. Lobino, R. Ramponi, and P. Laporta, E. Cianci and V. Foglietti
%%% "Observation of Dynamic Localization in Periodically CurvedWaveguide Arrays",
%  Phys. Rev. Lett. {\bf 96}, 243901(2006).

\end{references}
\end{document}